# Chemical-Potential Multiphase Lattice Boltzmann Method with Superlarge Density Ratios

Binghai Wen[1*], Liang Zhao[2], Wen Qiu[1], Yong Ye[1], Xiaowen Shan[3†]

[1]Guangxi Key Lab of Multi-Source Information Mining & Security, Guangxi Normal University, Guilin 541004, China

[2]College of Physical Science and Technology, Yangzhou University, Jiangsu 225009, China

[3]Department of Mechanics and Aerospace Engineering, Southern University of Science and Technology, Shenzhen, Guangdong 518055, China

**Abstract:** The liquid–gas density ratio is a key property of multiphase flow methods to model real fluid systems. Here, a chemical-potential multiphase lattice Boltzmann method is constructed to realize extremely large density ratios. The simulations show that the method reaches very low temperatures, at which the liquid–gas density ratio is more than $10^{14}$, while the thermodynamic consistency is still preserved. Decoupling the mesh space from the momentum space through a proportional coefficient, a smaller mesh step provides denser lattice nodes to exactly describe the transition region and the resulting dimensional transformation has no loss of accuracy. A compact finite-difference method is applied to calculate the discrete derivatives in the mesh space with high-order accuracy. These enhance the computational accuracy of the nonideal force and suppress the spurious currents to a very low level, even if the density ratio is up to tens of thousands. The simulation of drop splashing verifies that the present model is Galilean invariant for dynamic flow field. An upper limit of the chemical potential is used to reduce the influence of nonphysical factors and improve the stability.

[*†] Corresponding authors. Email: oceanwen@gxnu.edu.cn (B. Wen), shanxw@sustech.edu.cn (X. Shan).

# 1. Introduction

In the past three decades, the lattice Boltzmann method (LBM) has been developed as a powerful tool for modeling complex fluid systems [1,2], especially for multiphase and multicomponent flows [3,4]. Because the Boltzmann equation assumes that the particles are uncorrelated prior to the collisions, it cannot directly describe a nonideal effect or phase transition. The nonideal interaction between particles has to be calculated by extra methods. Several popular multiphase models have been developed by the LBM community, including the color-gradient model [5], pseudopotential model [6,7], free energy model [8,9], phase-field model [10], and entropic model [11]. Rigorous derivation of the Enskog equation has provided a unified framework to treat lattice Boltzmann models for nonideal fluids [12,13]. In the early stages, the above multiphase models were limited to small liquid–gas density ratios because of the numerical instability in the interface layer. Simulations of real liquids and gases often require high density ratios. In particular, water-vapor systems usually require a large density ratio of up to 1000.

Many studies have achieved a large density ratio by recovering the interface capturing equations. These studies use two sets of distribution functions, one for evolving the density field to solve the Navier–Stokes equations and the other for evolving the order parameter to solve the Cahn–Hilliard or Allen–Cahn equations [14,15]. Inamuro *et al*. [16] first achieved a large density ratio on a two-phase immiscible free-energy model together with the Cahn–Hilliard equation. Lee and Lin [17] achieved a large density ratio on a phase-field model with a stable discretization algorithm. Fakhari *et al*. [18,19] extended the model to high Reynolds number flows using a multiple-relaxation-time (MRT) collision operator and achieved mass conservation. Zhang *et al*. [20] recently compared the model with quadtree adaptive mesh refinement with an arbitrary Lagrangian Eulerian finite element method. Zheng *et al*. [14] proposed an improved model to accurately recover the Cahn–Hilliard equation and did not require pressure correction. He *et al*. [21] used the model to study the roles of the wettability and surface tension in droplet formation during inkjet printing. Yan and Zu [22] combined Inamuro's model [16] and the free energy model [23,24] to simulate multiphase flows with a large density ratio and partial wetting surfaces. Wang *et al*. [25]

developed a multiphase lattice Boltzmann flux solver with large density ratios, in which a stable weighted difference scheme was applied to solve the Cahn–Hilliard equation. The flux solver was then improved by modifying the calculation of the equilibrium density distribution function [26]. Liang *et al*. [15] recently proposed a large-density-ratio phase-field model by solving the conservative Allen–Cahn equation, and they extended it to the axisymmetric MRT method [27]. In the color-gradient model, Leclaire *et al*. [28] used an enhanced equilibrium distribution function to simulate immiscible multiphase flows and achieved a large density ratio. Ba *et al*. [29] applied a MRT collision operator to improve the stability of the color-gradient model and simulated steady flows with a large density ratio. Ridl and Wagner [30] proposed a framework for simulating multicomponent and multiphase systems and achieved a density ratio of 1700.

The pseudopotential model is the most popular model in the multiphase-flow LBM [7]. Conceptually, it originates from the long-range intermolecular interaction, and its mesoscopic interaction potential gives a nonideal gas equation of state (EOS) at the macroscopic level and the associated behaviors of the multiphase flow and phase transitions [31,32]. When an EOS is specified, not only a high density ratio, but also thermodynamic consistency is expected, which requires the liquid and gas coexistence densities to agree with the analytical predictions of the Maxwell equal-area construction. Shan [31] and Sbragaglia *et al*. [33] used high-order isotropic discrete gradient operators to suppress the spurious currents and increase the density ratio. Khajepor *et al*. [34] applied multi-pseudopotential interaction to achieve thermodynamic consistency. Yuan and Schaefer [35] developed an improved equation for the effective mass, which enabled the pseudopotential model to incorporate the common EOSs of real fluids. Wagner and Pooley [36] introduced a factor before van der Waals EOS. It widened the phase interface and obtained the large density ratio. Kupershtokh *et al*. [37] used a factor of the order of 0.01 before EOS and linearly combined Shan–Chen [6] and Zhang–Chen [32] nonideal forces by a weighted coefficient. Huang *et al*. [38] performed simulations with a large density ratio and analyzed the relationship of the interface width and the EOS parameters. Li *et al*. [39,40] proposed an improved forcing scheme to enhance the stability of the pseudopotential model with a large density ratio. Hu *et al*. [41,42] pointed out that although the coefficient did not influence the

Maxwell construction of EOSs, it made the EOSs different from their original versions. Qin *et al*. [43] introduced a high-order difference scheme to achieve a large density ratio. The simulation results showed that with the high-accuracy scheme, the nonideal forces independently evaluated by the Shan–Chen and Zhang–Chen schemes are the same, and thus their combination and the weight coefficient are completely unnecessary. This indicates that the numerical errors in the discrete algorithms play a more important role than expected.

When a multiphase flow system is modeled, the thermodynamic consistency and Galilean invariance should be satisfied as two fundamental requirements. The former ensures that the model correctly produces interphase equilibrium, while the latter is related to accurate description of the multiphase fluid motion. Based on free energy theory, Wen *et al*. [9] directly calculated the nonideal force by the thermodynamic pressure tensor and proposed a multiphase model that meets the requirements of thermodynamic consistency and Galilean invariance, which was verified by theoretical analyses and numerical simulations. Recently, Wen *et al*. [44] replaced the divergence of the pressure tensor by the gradient of the chemical potential and proposed a chemical-potential multiphase model. Because calculations of the pressure tensor and its divergence were avoided, the chemical-potential model had lower temporal and spatial complexities. A chemical-potential boundary condition was implemented to express the surface wettability, and the contact angle could then be linearly tuned by the chemical potential of the surface. Furthermore, a real-time scheme was designed for accurate measurement of the contact angle [45]. Because the scheme is based on the interfacial geometry near the three-phase contact line, the measurement reflects the microscopic contact angle and is suitable for capturing the dynamic contact angle with or without gravity. However, in the chemical-potential model, the liquid–gas density ratio is more than 100, while the spurious current is up to the order of 0.01, which limit its application in simulations of real multiphase systems [44].

In this paper, we propose a chemical-potential multiphase LBM with extremely large density ratios, thermodynamic consistency, and Galilean invariance. An optional proportional coefficient is introduced to decouple the computational mesh from the momentum space. Together with a high-accuracy compact finite-difference scheme, common

EOSs can achieve extremely large density ratios, while the thermodynamic consistency is still preserved and the spurious current is suppressed to a very low level.

## 2. Lattice Boltzmann method

LBM originated from the lattice gas automaton concept and kinetic theory [1,2,46]. The intrinsic mesoscopic properties make it outstanding in modeling complex fluid systems involving interfacial dynamics and phase transitions [4,7]. The lattice Boltzmann equation (LBE) is fully discretized in space, time, and velocity. Its single-relaxation-time (SRT) version can be concisely expressed as [47]

$$f_i(\boldsymbol{x}+\boldsymbol{e}_i\delta t, t+\delta t) - f_i(\boldsymbol{x},t) = -\frac{1}{\tau}[f_i(\boldsymbol{x},t) - f_i^{(eq)}(\boldsymbol{x},t)] + F_i, \qquad (1)$$

where $f_i(\boldsymbol{x},t)$ is the particle distribution function at time $t$ and lattice site $\boldsymbol{x}$, moving along the direction defined by the discrete velocity vector $\boldsymbol{e}_i$ with $i=0,...,N$, $\tau$ is the relaxation time, and $F_i$ is the body force term. $f_i^{(eq)}(\boldsymbol{x},t)$ is the equilibrium distribution function

$$f_i^{(eq)}(\boldsymbol{x},t) = \omega_i \rho(\boldsymbol{x},t)\left[1 + \frac{(\boldsymbol{e}_i\cdot\boldsymbol{u})}{c_s^2} + \frac{(\boldsymbol{e}_i\cdot\boldsymbol{u})^2}{2c_s^4} - \frac{(\boldsymbol{u})^2}{2c_s^2}\right], \qquad (2)$$

where $\boldsymbol{u}$ is the fluid velocity. For the two-dimensional nine-velocity (D2Q9) model on a square lattice, the weighting coefficients are $\omega_0 = 4/9$, $\omega_{1-4} = 1/9$, and $\omega_{5-8} = 1/36$. The sound speed is $c_s = c/\sqrt{3}$, where $c = \delta x/\delta t$, in which $\delta x$ and $\delta t$ are the space step and time step, respectively. The discrete velocity set is given by $\boldsymbol{e} = c[(0,0),(1,0),(0,1),(-1,0),(0,-1),(1,1),(-1,1),(-1,-1),(1,-1)]$. $\tau$ is related to the kinematic viscosity by $\nu = c_s^2(\tau - 0.5\delta t)$. The fluid density and momentum at a lattice node can be defined by

$$\rho = \sum_{i=0}^{8} f_i, \qquad \rho\boldsymbol{u} = \sum_{i=0}^{8} \boldsymbol{e}_i f_i. \qquad (3)$$

$f_i$ and $e_i f_i$ can be considered as a mass component of a lattice node and the corresponding momentum component [48].

The MRT version of LBE improves the numerical stability and computational accuracy, and it can be expressed as [49]

$$f_i(\boldsymbol{x}+\boldsymbol{e}_i\delta t, t+\delta t) - f_i(\boldsymbol{x},t) = -\mathbf{M}^{-1}\cdot\mathbf{S}\cdot[\mathbf{m}-\mathbf{m}^{(eq)}] + F_i, \qquad (4)$$

where $\mathbf{m}$ and $\mathbf{m}^{(eq)}$ represent the velocity moments of the distribution functions and their equilibria, respectively, $\mathbf{M}$ is a transformation matrix which linearly transforms the distribution functions to the velocity moments, $\mathbf{m}=\mathbf{M}\cdot\mathbf{f}$, and $\mathbf{f}=\mathbf{M}^{-1}\cdot\mathbf{m}$, where $\mathbf{f}=(f_0, f_1, ..., f_8)$ for the D2Q9 model. $\mathbf{S}$ is a diagonal matrix of nonnegative relaxation times: $\mathbf{S}=\text{diag}(0, s_e, s_\varepsilon, 0, s_q, 0, s_q, s_v, s_v)$. In this paper, the relaxation times are given by $s_e=1.64$, $s_\varepsilon=1.54$, $s_q=1.9$, and $s_v=1/\tau$ for the simulations with the MRT LBE.

The external force is brought into effect through forcing technology [50,51]. We chose the exact difference method proposed by Kupershtokh *et al.* to incorporate the nonideal force $\boldsymbol{F}$ into LBE [37]:

$$F_i = f_i^{(eq)}(\rho, \boldsymbol{u}+\delta\boldsymbol{u}) - f_i^{(eq)}(\rho, \boldsymbol{u}), \qquad (5)$$

where $\delta\boldsymbol{u}=\delta t\boldsymbol{F}/\rho$. The body force term $F_i$ is simply equal to the difference of the equilibrium distribution functions before and after the nonideal force acting on the fluid during a time step. Correspondingly, the macroscopic fluid velocity is redefined as the average momentum before and after the collision: $\boldsymbol{v}=\boldsymbol{u}+\delta t\boldsymbol{F}/(2\rho)$.

## 3. Chemical-potential multiphase model

### 3.1 Nonideal force evaluation by the chemical potential

The chemical potential is the partial molar Gibbs free energy at constant pressure [52]. Both the Onsager and Stefan–Maxwell formulations of irreversible thermodynamics recognize that the chemical potential gradient is the driving force for isothermal mass transport [53]. Movement of molecules from higher to lower chemical potential is accompanied by a release of free energy, and the chemical or phase equilibrium is achieved at the minimum free energy. Following the classical capillarity theory of van der Waals, the free energy functional within a gradient-squared approximation is [8,9,54]

$$\Psi = \int [\psi(\rho) + \frac{\kappa}{2}|\nabla\rho|^2] d\boldsymbol{x},  \quad (6)$$

where $\rho$ is the local density, $\kappa$ is the surface tension coefficient, $\psi$ is the bulk free energy density at a given temperature, and the square of the gradient term gives the free energy contribution from density gradients in an inhomogeneous system. The chemical potential can be derived from the free energy density functional [14,52,54]:

$$\mu = \psi'(\rho) - \kappa\nabla^2\rho. \quad (7)$$

The nonlocal pressure is related to free energy by

$$p = p_0 - \kappa\rho\nabla^2\rho - \frac{\kappa}{2}|\nabla\rho|^2, \quad (8)$$

with the general EOS defined by the free energy density

$$p_0 = \rho\psi'(\rho) - \psi(\rho). \quad (9)$$

The thermodynamic pressure tensor of nonuniform fluids contains the nondiagonal terms

$$P_{\alpha\beta} = p\delta_{\alpha\beta} + \kappa\partial_\alpha\rho\partial_\beta\rho. \quad (10)$$

where $\delta_{\alpha\beta}$ is the Kronecker delta function. With respect to the ideal gas, the excess pressure can be directly calculated by [9,55]

$$\boldsymbol{F} = -\nabla\cdot\vec{\vec{P}} + \nabla\cdot\vec{\vec{P}}_0, \quad (11)$$

where $\vec{\vec{P}}_0 = c_s^2\rho\vec{\vec{I}}$ is the ideal-gas EOS, in which $\vec{\vec{I}}$ is the unit tensor. Substituting Eqs. (7) and (9) into the divergence of the pressure tensor, an elegant relationship can be obtained to relate the gradient of the chemical potential: $\nabla\cdot\vec{\vec{P}} = \rho\nabla\mu$. Thus, the nonideal force can be evaluated by the chemical potential [44]:

$$\boldsymbol{F} = -\rho\nabla\mu + c_s^2\nabla\rho. \quad (12)$$

Because the derivation is within thermodynamics, it is expected that the phase transition induced by the nonideal force theoretically satisfies the thermodynamic consistency and

Galilean invariance, which have been confirmed by numerical simulations of static and dynamic fluids [9,44].

### 3.2 Equations of state and chemical potentials

Through Eq. (7) and (9), the bulk free energy density connects an EOS and its chemical potential. The general solution of the bulk free energy density can be determined by solving Eq. (9), which is a typical linear ordinary differential equation:

$$\psi = \rho(\int \frac{p_0}{\rho^2} d\rho + C), \tag{13}$$

where $C$ is a constant and does not appear in the nonideal force evaluation. Substituting Eq. (13) with a specific EOS into Eq. (7), we can then obtain the relevant chemical potential.

The chemical potentials of some widely used EOSs have been solved [44]. The VDW EOS is the most famous cubic EOS:

$$p_0 = \frac{\rho RT}{1-b\rho} - a\rho^2, \tag{14}$$

where $R$ is the universal gas constant, $a$ is the attraction parameter, and $b$ is the volume correction. The chemical potential is

$$\mu = RT[\ln(\frac{\rho}{1-b\rho}) + \frac{1}{1-b\rho}] - 2a\rho - \kappa \nabla^2 \rho. \tag{15}$$

The term with the Laplace operator gives the chemical potential contribution from the density gradients. The rest of the right part comes from the derivative of the bulk free energy density and it corresponds to the EOS. The Redlich–Kwong (RK) EOS is generally more accurate than the VDW EOS due to the modification of the attraction term:

$$p_0 = \frac{\rho RT}{1-b\rho} - \frac{a\alpha(T)\rho^2}{(1+b\rho)}, \tag{16}$$

where $\alpha(T) = 1/\sqrt{T}$. The Soave modification (RKS) has a more complicated temperature function: $\alpha(T) = [1+(0.480+1.574\omega-0.176\omega^2)(1-\sqrt{T_r})]^2$, where $\omega$ is the acentric factor. Both equations share the same chemical potential

$$\mu = RT \ln \frac{\rho}{1-b\rho} + \frac{RT}{1-b\rho} - \frac{a\alpha(T)}{b} \ln(1+b\rho) - \frac{a\alpha(T)\rho}{1+b\rho} - \kappa \nabla^2 \rho. \tag{17}$$

The Peng–Robinson (PR) EOS is often superior in predicting liquid densities:

$$p_0 = \frac{\rho RT}{1-b\rho} - \frac{a\alpha(T)\rho^2}{1+2b\rho - b^2\rho^2}, \tag{18}$$

where the temperature function is $\alpha(T) = [1 + (0.37464 + 1.54226\omega - 0.26992\omega^2) \times (1-\sqrt{T/T_c})]^2$. Its chemical potential is

$$\mu = RT \ln \frac{\rho}{1-b\rho} - \frac{a\alpha(T)}{2\sqrt{2}b} \ln \frac{\sqrt{2}-1+b\rho}{\sqrt{2}+1-b\rho} + \frac{RT}{1-b\rho} - \frac{a\alpha(T)\rho}{1+2b\rho-b^2\rho^2} - \kappa \nabla^2 \rho. \tag{19}$$

The Carnahan–Starling (CS) EOS tends to give better approximation of the repulsive term:

$$p_0 = \rho RT \frac{1+b\rho/4+(b\rho/4)^2-(b\rho/4)^3}{(1-b\rho/4)^3} - a\rho^2, \tag{20}$$

where the chemical potential is

$$\mu = RT[\frac{3-b\rho/4}{(1-b\rho/4)^3} + \ln \rho + 1] - 2a\rho - \kappa \nabla^2 \rho. \tag{21}$$

In this study, the attraction parameter and volume correction are $a=9/49$ and $b=2/21$ for the VDW EOS, $a=2/49$ and $b=2/21$ for the PR and RKS EOSs, and $a=1$ and $b=4$ for the CS EOS. The universal gas constant is $R=1$. The acentric factor $\omega$ is 0.344 for water and 0.011 for methane. To relate the numerical results to real physical properties, we define the reduced variables $T_r = T/T_c$ and $\rho_r = \rho/\rho_c$, where $T_c$ is the critical temperature and $\rho_c$ is the critical density. Unless otherwise stated, the following temperature and density refer to the reduced temperature and reduced density, respectively.

### 3.3 Analytical solution of the density profile

Because the free energy density is already in the chemical potential model [44], it is convenient to solve the density and gradient distributions of the transition region. Let us consider an isothermal liquid–gas system which has a planar phase interface and develops along the $y$ coordinate. In the equilibrium state, the domain has the boundary conditions $\rho(-\infty) = \rho_g$ and $\rho(+\infty) = \rho_l$, where $\rho_g$ and $\rho_l$ are the bulk densities of the gas and liquid phases, respectively, and $(\rho_g + \rho_l)/2$ is the origin of the $y$ coordinate. The mechanical equilibrium condition $\nabla \cdot \vec{P}(x) = 0$ can be ensured by the equilibrium of the chemical potential $\mu(x) = \mu_b$, where $\mu_b$ is the bulk chemical potential. Solving Eq. (7) in the one-dimensional system gives

$$\frac{\kappa}{2}(\frac{d\rho}{dy})^2 = \psi(\rho) - \psi(\rho_b) - \mu_b(\rho - \rho_b), \tag{22}$$

where $\rho_b$ is equal to $\rho_g$ or $\rho_l$. Transforming Eq. (22) and integrating gives

$$\pm \int_0^y dy = \int_{\rho_0}^{\rho} \frac{d\rho}{\sqrt{\frac{2}{\kappa}[\psi(\rho) - \psi(\rho_b) - \mu_b(\rho - \rho_b)]}}, \tag{23}$$

where the sign of the left part is negative for $\rho_b = \rho_g$ in the gas phase or positive for $\rho_b = \rho_l$ in the liquid phase. For a given density $\rho \in [\rho_g, \rho_l]$, integrating Eq. (23) gives the corresponding $y$ value, and the density distribution with $y$ is then obtained, which is shown as the analysis solution in Fig. 1. In turn, we can calculate the density gradient distribution with $y$ by

$$\frac{d\rho}{dy} = \pm \sqrt{\frac{2}{\kappa}[\psi(\rho) - \psi(\rho_b) - \mu_b(\rho - \rho_b)]} \ . \tag{24}$$

This is shown as the analysis solution in Fig. 3. The figures show that both distributions of the density and density gradient are not symmetrical about the origin. Therefore, the profile of the transition region is different from a hyperbolic tangent function, of which the distributions are symmetrical about the origin [43].

## 4. Decoupling the computational mesh from the momentum space

LBE discretizes time and phase space of which configuration space is of a lattice structure and momentum space is reduced to a small set of discrete momenta [56,57]. In practice, the discretization of configuration space is usually coupled with that of momentum space. Typically, the nonzero $|e_{ix}|$ and $|e_{iy}|$ are equal to $\delta x$, which determines a simple square lattice, such as that of the D2Q9 model. Although the configuration simplifies physical analyses, program codes, and numerical computations, it inhibits the use of a nonuniform grid. He *et al.* proposed a nonuniform grid algorithm, in which the computational mesh is uncoupled from the discretization of momentum space and can have an arbitrary shape [58]. However, an interpolation step has to be applied to supplement the missing distribution functions on the computational mesh. Cao *et al.* [59] suggested that the lattice symmetry is not essential for recovering the macroscopic equations, and besides using a nonuniform grid, the number of lattice links could be different from that of the particle distribution functions. Conceptually, LBM can be divided into three layers: the physical space, the momentum space, and the mesh space (computational mesh). It is a dimensional transformation that transforms a quantity between the physical space and the momentum space. Now, we attempt to decouple the computational mesh from the momentum space, and the mesh space then has an independent step length. Mathematically, the constraint between the momentum and mesh spaces can be moderately relaxed using an optional proportion between the lattice step and mesh step, which can be seen as the length units in the momentum and mesh spaces, respectively. This proportional relationship is much better than an interpolation algorithm, and it determines dimensional transformations that transform quantities between the momentum and mesh spaces. Notably, there is no loss of accuracy in the transformation process.

### 4.1 Dimensional transformation

Let us use an optional proportional coefficient $k$ to connect the lattice step $\delta x$ of the momentum space with the mesh step $\tilde{\delta x}$ of the computational mesh:

$$\tilde{\delta x} = k \delta x. \tag{25}$$

Here, if the dimension of a quantity contains a unit of length, the symbol of the quantity in the mesh space is marked with a superscript, such as for the length, velocity, and pressure. The time, density, and temperature are considered to be independent of the length, so they have the same values and symbols in the two spaces. It should be noted that in the context of LBM, the local density is the summation of the distribution functions on a lattice site, and it is independent of the length. In contrast, the mass is calculated by the spatial integral of the local density, and it is dependent on the length. Because $\delta \hat{x}/\delta t = k \delta x/\delta t$, the lattice constant is $\hat{c} = kc$, and then the transformations of the velocities are simple:

$$\hat{\bm{e}}_i = k\bm{e}_i, \quad \hat{\bm{u}} = k\bm{u}, \quad \hat{c}_s = kc_s. \tag{26}$$

The spatial derivative operations are also influenced by the proportion. The gradient and Laplace operators contribute negative first-power and second-power of the proportional coefficient, that is,

$$\hat{\nabla} = k^{-1}\nabla, \quad \hat{\nabla}^2 = k^{-2}\nabla^2. \tag{27}$$

Dimensional analysis is used to determine the coefficients of some quantities with more complex dimensions. The free energy density is the ratio of the energy to the volume. Its dimension can be expressed as $MV^2/L^3$, where $M$, $L$, and $V$ represent the mass, length, and velocity dimensions, respectively. In the LBM context, the dimension of the free energy density can be expressed as $\rho_0 L^2/T_0^2$, where $T_0$ and $\rho_0$ are the time and density dimensions. Thus, the transformations of the free energy density can be expressed as

$$\hat{\psi}(\rho) = k^2 \psi(\rho). \tag{28}$$

Equations (8) and (9) indicate that the dimension of the pressure is the same as that of the free energy density. Therefore, the pressures have the transformations

$$\hat{p}_0 = k^2 p_0 \tag{29}$$

and

$$\hat{p} = k^2 p. \tag{30}$$

Equation (7) indicates that the chemical potential and derivative of the free energy density have the same dimension, which can be simplified as $L^2/T_0^2$. The chemical potential has the transformation

$$\mu = k^2 \hat{\mu}. \tag{31}$$

The nonideal force has the same dimension as the product of the density and the gradient of the chemical potential according to Eq. (12), and it can be expressed as $\rho_0 L/T_0^2$. The nonideal force has the transformation

$$\boldsymbol{F} = k\hat{\boldsymbol{F}}. \tag{32}$$

In the present model, the evolution of LBE is in the mesh space. Because the universal gas constant, attraction parameter, and volume correction have complex dimensions, to simplify the analysis, we limit the equations involving these parameters, such as EOS and free energy density, are calculated in the momentum space. Therefore, the chemical potential in the mesh space is obtained by

$$\hat{\mu} = k^2 \psi'(\rho) - \hat{\kappa}\hat{\nabla}^2 \rho, \tag{33}$$

where $\psi'(\rho)$ represents the right part of the chemical potential without Laplace operator term. It is calculated in the momentum space and transformed into the mesh space by the coefficient $k^2$.

The coefficient $k^2$ in Eq. (29) is similar to the factor used before EOSs in the literature [36,37,41], which widen the phase interface like the studies [38,40]. However, the factor in the literature makes the EOSs different from their original versions [41,42]; in contrast, the coefficients used in the above equations neither influence the Maxwell equal-area construction nor change the meanings of the equations, such as EOS, pressure, chemical potential and nonideal force. The proportional coefficient defines the dimensional transformation between the momentum and mesh spaces. Because the transformation has no loss of accuracy, the present chemical-potential multiphase model is mathematically equivalent to the previous versions [9,44]. Thus, the present model theoretically satisfies the thermodynamic consistency and Galilean invariance, which are numerically confirmed by the simulations reported in Section 5.

**4.2 Gentle profile of the transition region**

If the proportional coefficient is less than one then the mesh step is less than the lattice step of the momentum space and describes the transition region in more detail. That is, the steep transition region in the momentum space is stretched into a gentle curve in the mesh space. In contrast, the density remains the same in both spaces, because the density unit is unchanged in both spaces.

The density profiles of the liquid–gas transition region in the momentum space and the various mesh spaces that connect to the momentum space through different proportional coefficients are shown in Fig. 1. We select the VDW and PR EOSs at the temperature of 0.6. The gray lines are the analysis solutions. With decreasing $k$, an increasing number of points are involved in the transition region. Taking the VDW EOS as an example, three points outline the steep transition region in the momentum space ($k=1$), whereas in the mesh space with $k=0.2$, there are eleven points to describe the same region and the curve is much gentler.

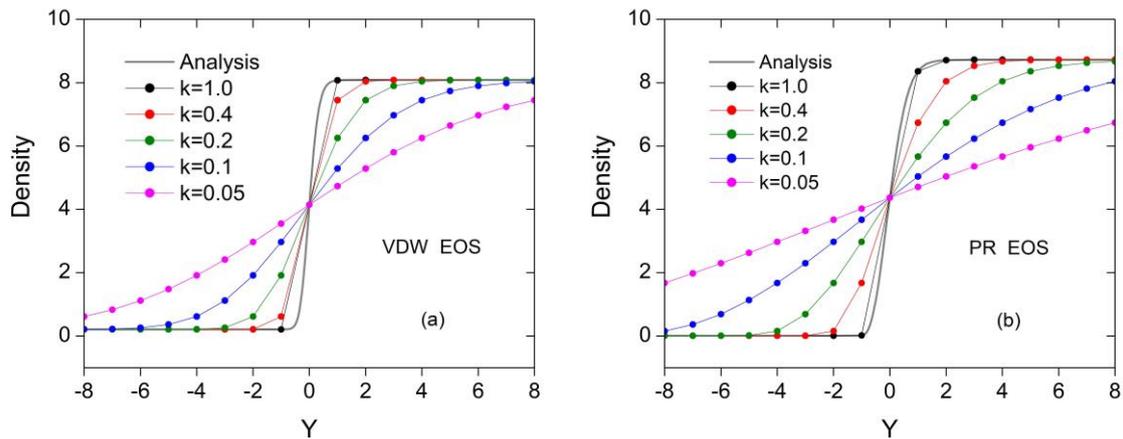

*Fig. 1. Density profiles of the liquid–gas transition region depicted as discrete nodes in the momentum space and the various mesh spaces defined by different proportional coefficients. It is clear that the density profiles become increasingly gentle with decreasing proportional coefficient.*

The errors in numerical calculations of multiphase simulations are mainly dependent on the accuracy of the discrete gradient calculations. Denser points make the gradient calculation more accurate. The relative errors of the gradient calculations determined by the central difference method (CDM) with second-order accuracy, which is the widely used in multiphase LBMs, are shown in Fig. 2 (open stars). The relative errors are very high in the momentum space ($k=1$). Although the relative errors decrease with decreasing $k$, they are not sufficiently good yet. In the next section, we will introduce a compact finite-difference method (CFM) to further improve the accuracy of the discrete gradient calculations.

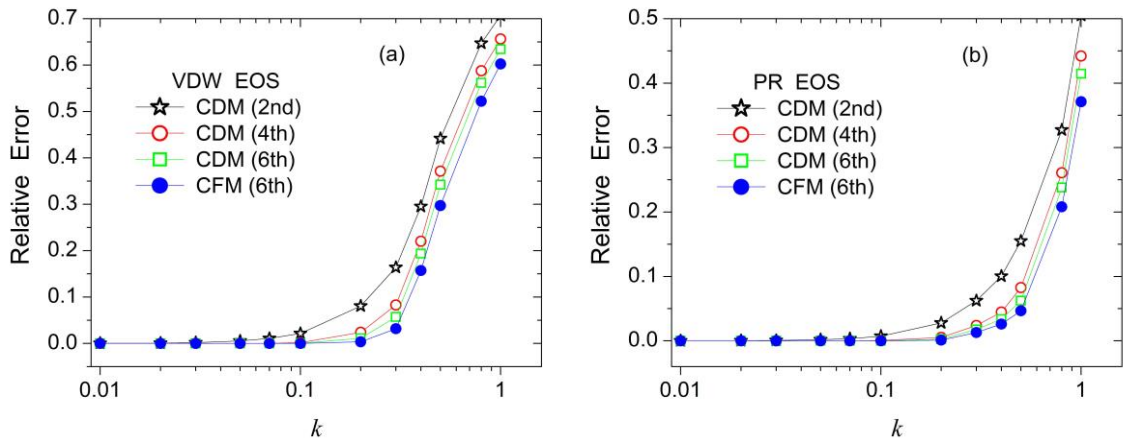

*Fig. 2. Relative errors of the gradient calculations determined by the second-, fourth-, and sixth-order CDMs and the sixth-order CFM plotted against the proportional coefficient $k$.*

### 4.3 High-accuracy calculation of the gradients

In numerical simulations of multiphase flows by LBM, it is popular to apply the second-order CDM to calculate the gradients of some characteristic quantities, such as the density, effective mass, and chemical potential. Inside the bulk phase, because the fluctuations of the characteristic quantities are very small, the results from the second-order CDM are acceptable. However, in the transition region between the gas and liquid phases, the changes of the characteristic quantities are violent. For example, the liquid–gas density ratio in the natural environment is more than 1000, but it is only two or three lattices to span the large density difference in simulations. This results in a rather steep phase interface, which is similar to a hyperbolic tangent function. When calculating the gradients in the transition regions, the second-order CDM causes considerable errors, and it even makes mathematically equivalent equations behave like different algorithms [43]. Therefore, it is necessary to use a more accurate method to calculate the derivatives and gradients in multiphase simulations.

The derivatives in a multiphase model can be calculated by several different difference schemes. Let $u(x)$ be a continuous and derivable function defined on the closed interval $[x_0, x_n]$. The interval is divided into $n$ subintervals on average: $x_i = x_0 + ih$, $h = (x_n - x_0)/n$, $i = 0, 1, 2, \cdots, n$. The common CDMs with increasing formal accuracy can be written as

$$\text{second order:} \quad u'_i = \frac{u_{i+1} - u_{i-1}}{2h} \tag{34}$$

$$\text{fourth order:} \quad u'_i = \frac{4}{3}\frac{u_{i+1} - u_{i-1}}{2h} - \frac{1}{3}\frac{u_{i+2} - u_{i-2}}{4h} \tag{35}$$

$$\text{sixth order:} \quad u'_i = \frac{3}{2}\frac{u_{i+1} - u_{i-1}}{2h} - \frac{3}{5}\frac{u_{i+2} - u_{i-2}}{4h} + \frac{1}{10}\frac{u_{i+3} - u_{i-3}}{6h}. \tag{36}$$

To obtain higher accuracy, we introduce CFM to calculate the derivatives in the present model. The three derivatives near node $i$ are related to the two differences by the undetermined coefficients method [60]:

$$\alpha u'_{i-1} + u'_i + \alpha u'_{i+1} = a\frac{u_{i+1} - u_{i-1}}{2h} + b\frac{u_{i+2} - u_{i-2}}{4h}. \tag{37}$$

The relations between the coefficients $a$, $b$ and $\alpha$ are derived by matching the Taylor series coefficients of various orders. This gives an $\alpha$-family of tridiagonal schemes with

$$a = \frac{2}{3}(\alpha + 2), \quad b = \frac{1}{3}(4\alpha - 1), \tag{38}$$

where $\alpha = 1/3$ results in a typical tridiagonal equation with formally sixth-order accuracy [60]:

$$\frac{1}{3}u'_{i-1} + u'_i + \frac{1}{3}u'_{i+1} = \frac{14}{9}\frac{u_{i+1} - u_{i-1}}{2h} + \frac{1}{9}\frac{u_{i+2} - u_{i-2}}{4h}. \tag{39}$$

The equation is usually solved by the Thomas algorithm [61], which can calculate all of the derivatives on a line at a time. In the present multiphase model, the gradients of the density and chemical potential are calculated by CFM. The second-order derivatives of the density are obtained by simply applying CFM twice. Because the scheme is highly accurate, we directly use the horizontal and vertical derivatives in the numerical simulations. The weighted combination of the derivatives of multiple directions, which is commonly used for CDM, is unnecessary.

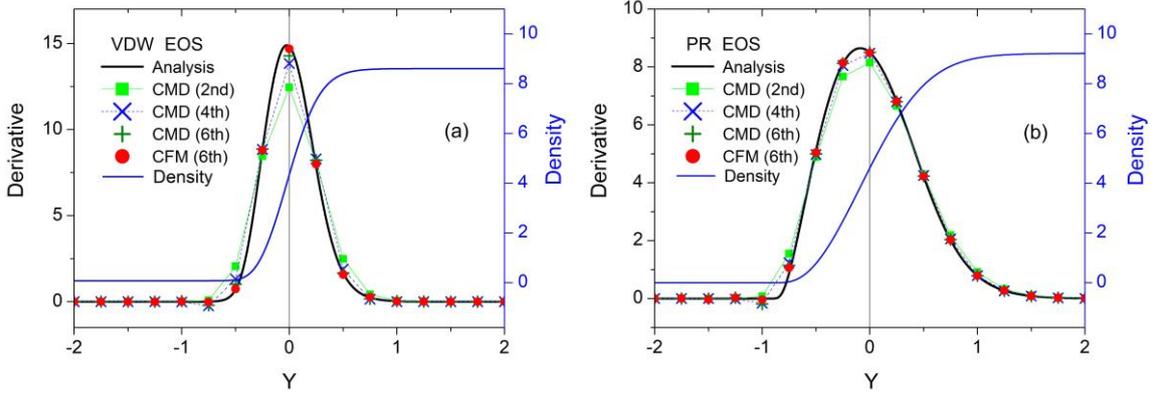

Fig. 3. Comparison of the numerical derivatives calculated by the second-, fourth-, and sixth-order CDMs and the sixth-order CFM.

Relative errors of the gradient calculations against the proportional coefficient $k$ at the temperature 0.6 are presented in Fig. 2 for several difference methods. It is clear that the high-order CDM is more accurate than the low-order one. However, the six-order CFM gives lower errors than the six-order CDM, although they have same-order formal accuracy. The density derivatives calculated by CDMs and CFM on the density profile are compared with the analytical results in Fig. 3. The VDW and PR EOSs are calculated at the temperature 0.5, and the proportional coefficient takes 0.25. The derivatives from the second-order CDM clearly deviate from the analytical solution. Essentially, the derivative calculated by the second-order CDM is just the average on the adjoining nodes without the local node, and the obvious deviations in the steep interface are almost unavoidable. The high-order difference methods use multiple adjacent nodes to calculate the derivative. The fourth-order and sixth-order CDMs improve the calculations, but the results from the sixth-order CFM shows the best agreement with the benchmarks. Therefore, we select the sixth-order CFM in the following simulations.

## 5. Simulations and discussion

The gentler transition region described by denser lattice nodes and the exact gradient calculations promote the nonideal force evaluation to obtain high accuracy. A series of numerical simulations involving first-order phase transitions are performed to demonstrate the qualities of the present multiphase method, including the thermodynamic consistency, Galilean invariance, density ratio, surface tension, spurious current, and stability. The D2Q9 model is used for the simulations. Unless otherwise stated, in the following sections, the proportional coefficient is $k = 0.1$ and the acentric factor of the PR EOS is $\omega = 0.344$ for water.

### 5.1. Superlarge density ratio

High-accuracy nonideal force evaluations enable the present multiphase method to simulate liquid–gas systems with very large density ratios. A one-dimensional liquid–gas system with two planar phase interfaces is used to calculate the two-phase coexistence densities in the equilibrium states. The theoretical values are obtained by solving the Maxwell equal-area construction equation and used as the benchmarks to verify the thermodynamic consistency. The height of the computational domain is 400 lattice units, while the width is optional. Periodic boundary condition is applied to the computational domain. With horizontal phase interfaces, the middle part of the domain is initialized as liquid, while the remaining part is initialized as gas. Four EOSs, namely, the VDW, RKS, PR, and CS EOSs, are evaluated and compared by the theoretical predictions. The calculation is performed for 200,000 time steps for each case. For comparison, the proportional coefficient and compact difference scheme are applied in the pseudopotential model, namely the Shan–Chen (SC) model [6]. The coefficient is used before EOS, as described in Eq. (29). The effective mass is evaluated based on EOS and the nonideal force is incorporated into LBE by the exact difference method [35,43].

The two-phase coexistence densities are shown in Fig. 4. The numerical results of the chemical-potential and pseudopotential models are all in excellent agreement with the benchmarks. To more clearly show the agreement, we select the logarithmic coordinate for the gas phase and the linear coordinate for the liquid phase. This verifies that with exact gradient calculation, both the chemical-potential and pseudopotential models can completely preserve the thermodynamic consistency. Nevertheless, the chemical-potential model reaches lower temperature and larger liquid–gas density ratio than the pseudopotential model. This is because of the difference between the pressure tensor derived from the pseudopotential model [38] and the thermodynamic pressure tensor, which is fully preserved in the chemical-potential model (i.e., Eq. (10)). The density ratios of the liquid phase to gas phase reach more than $5.5 \times 10^{10}$ for the VDW EOS, $6.9 \times 10^{11}$ for the PR EOS, $2.8 \times 10^{13}$ for the RKS EOS, and $3.4 \times 10^{14}$ for the CS EOS. We calculated the multiphase system on the SRT and MRT models with various relaxation times. The results of the MRT model with $\tau = 0.8$ are almost the same as those of the SRT model with $\tau = 3.0$, so the cases are independent of the relaxation time. A comprehensive comparison of the pseudopotential model and the chemical-potential model is in preparation. The distributions of the density and chemical potential calculated by PR EOS at the temperature 0.5 are shown in Fig. 5. To make the figure clearer, the height of the computational domain is changed to 600 lattice units. The numerical results are consistent to the theoretical analysis that the chemical potential should be constant in the bulk phases in the equilibrium state. The waves of the chemical potential at the liquid–gas transition regions result in the nonideal force and the phase separation.

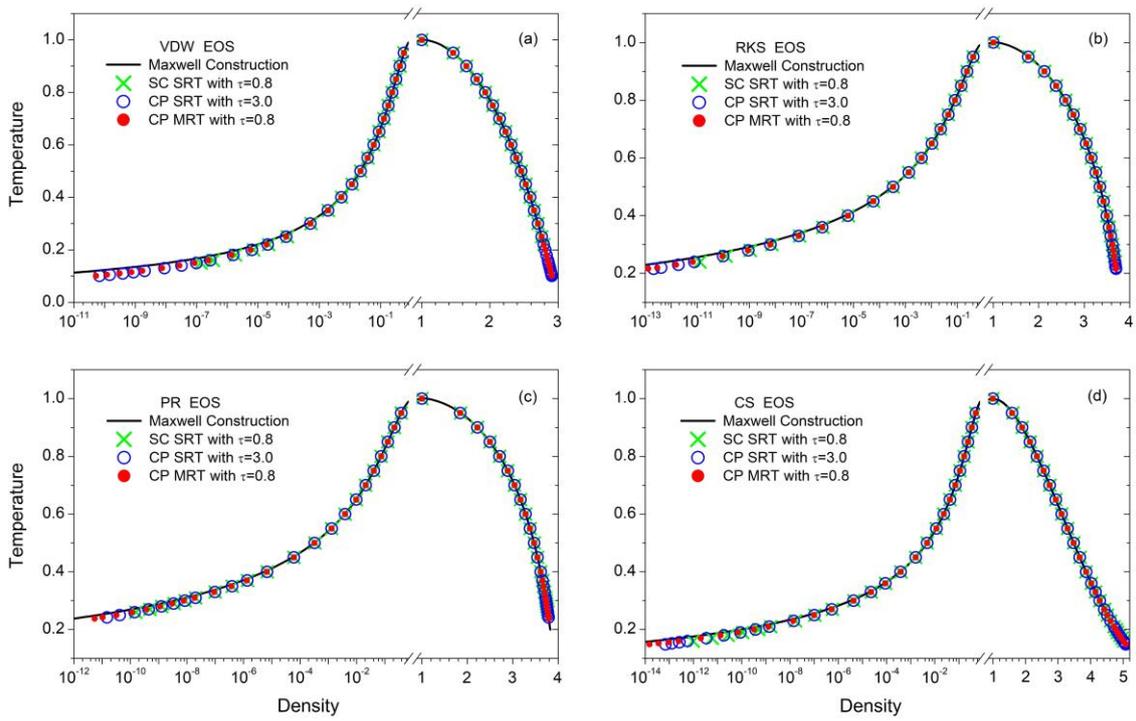

*Fig. 4. Two-phase coexistence densities calculated by the (a) VDW, (b) RKS, (c) PR, and (d) CS EOSs on the chemical-potential (CP) model and pseudopotential (SC) model compared with the theoretical predictions calculated by the Maxwell equal-area construction.*

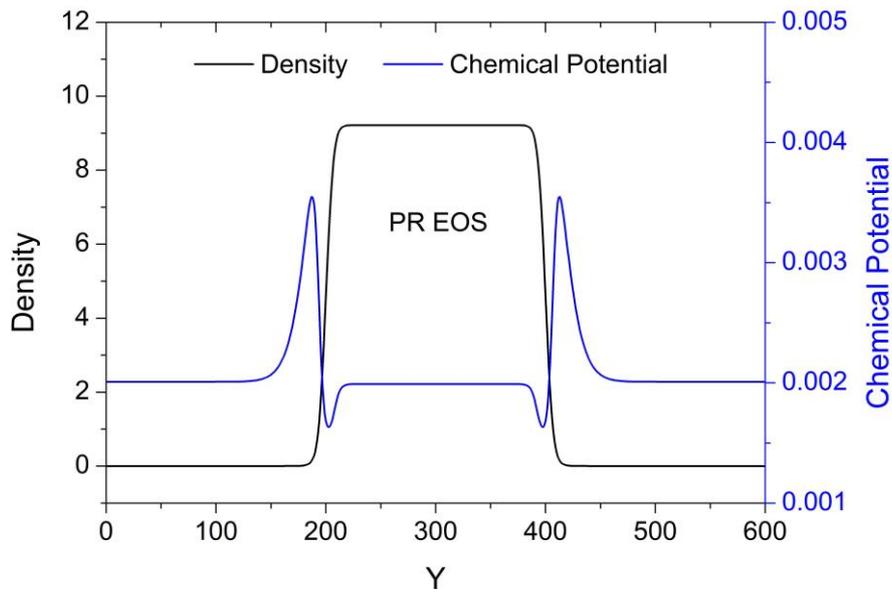

*Fig. 5. The distributions of density and chemical potential calculated by PR EOS.*

## 5.2. Interface width and surface tension

Theoretically, the proportional coefficient does not influence the interface width and the surface tension in the momentum space. That is, the values calculated in mesh spaces with different proportional coefficients should be the same when they are transformed into the momentum space. A one-dimensional system like that in the above section is used. The height of the computational domain is 800 lattice units to contain wider interfaces. The middle part of the domain is initialized as liquid, while the remaining part is initialized as gas. The temperature is $T_r = 0.6$. The proportional coefficient gradually increases from 0.05 to 0.3. Six EOSs are used, of which the PRM indicates the PR EOS with acentric factor $\omega = 0.011$ for methane.

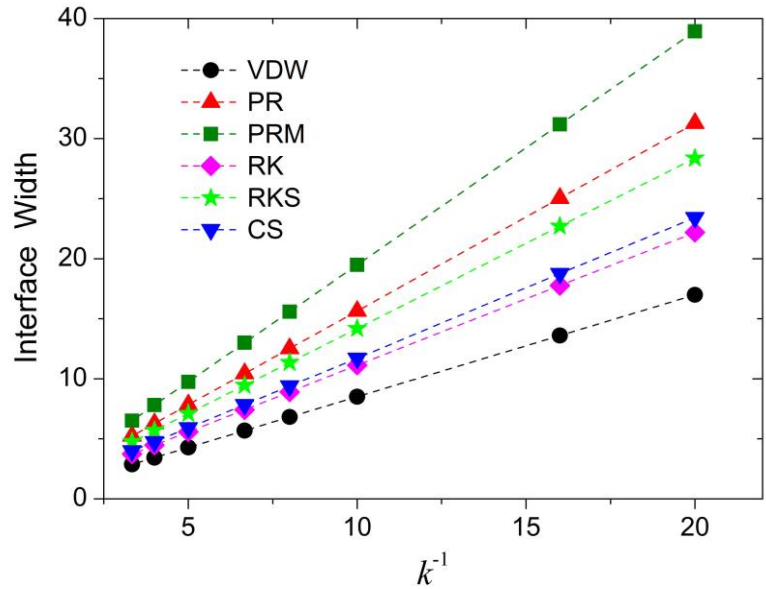

*Fig. 6. Interface width plotted against the inverse of the proportional coefficient. The slopes are equal to the corresponding widths in the momentum space.*

For the same EOS and temperature, when the proportional coefficient is less than one, the transition region widens. The interface widths are inversely proportional to the coefficient, as shown in Fig. 6. In other words, they have the same width when they are transformed into the momentum space through Eq. (25), which is equal to the slope. The mesh nodes that model a physical system in the mesh space are $k^{-D}$ times that of the momentum space, where $D$ is the spatial dimension. The surface tension in the mesh space also changed with the proportional coefficient. Nevertheless, the surface tension values calculated by each EOS are highly consistent when they are transformed into the momentum space as shown in Fig. 7. This verified that the transformations between the momentum space and computational mesh are stable and accurate.

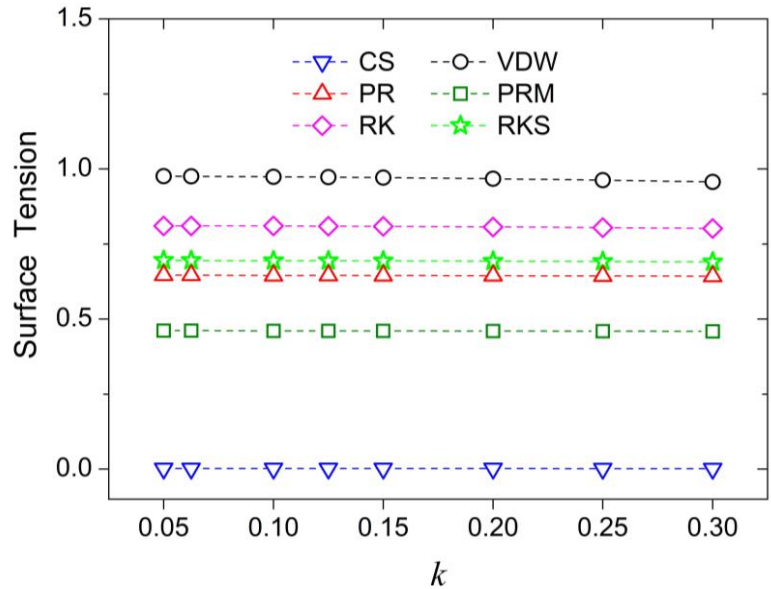

*Fig. 7. Surface tension calculated in the different mesh spaces. The surface tension values are highly consistent when they are transformed into the momentum space.*

### 5.3. Young–Laplace equation

The surface tension is a fundamental physical property in research of capillary phenomena and surface wettability. On a curved interface, the surface tension can be described by the Young–Laplace equation, which relates the capillary pressure difference sustained across the interface:

$$\Delta P = \sigma/R, \tag{40}$$

where $\sigma$ is the surface tension and $\Delta P$ is the pressure difference between inside and outside a curved interface with radius of curvature $R$. Following capillary theory [54,55], the surface tension of the liquid–gas interface can also be defined as the integral of the mismatch between the normal and transverse components of the pressure tensor along the coordinate normal to the interface. Here, the integral passes through the drop center along the vertical direction.

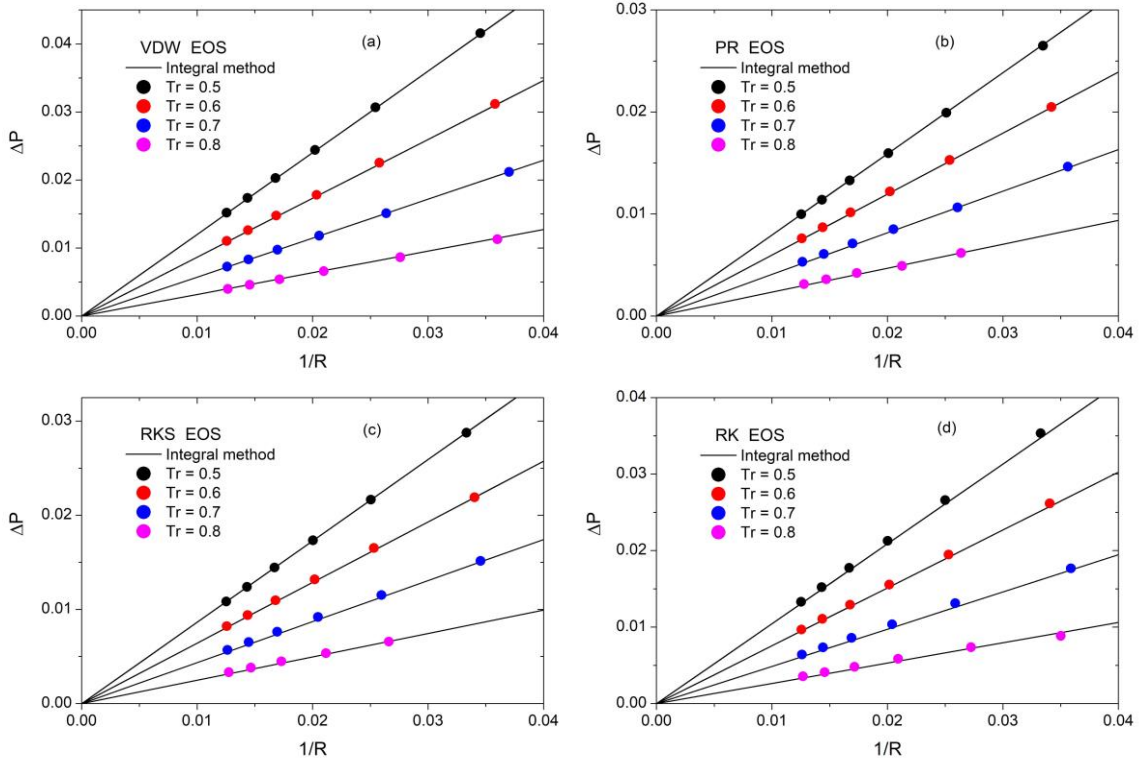

*Fig. 8. Series of static drops surrounded by gas in a gravity-free environment simulated to confirm the Young–Laplace equation. The slopes of the lines through the colored spots are the surface tension values, which are in excellent agreement with those calculated by the integral method.*

The computational domain is a square flow field with a side length of 300 lattice units. The density field is initialized as follows [38]:

$$\rho(x,y) = \frac{\rho_g + \rho_l}{2} + \frac{\rho_g - \rho_l}{2}\tanh\left[\frac{2(r-r_0)}{W}\right], \tag{41}$$

where $\rho_g$ and $\rho_l$ are the two-phase coexistence densities obtained by the Maxwell equal-area construction, $W = 10$ is the initial interface width, $r = \sqrt{(x-x_0)^2 + (y-y_0)^2}$, where $(x_0, y_0)$ is the center position of the domain, and $r_0$ is the initial drop radius, which changes from 30 to 80 lattice units. The relaxation time is $\tau = 1.5$ and the temperatures are chosen to be 0.5, 0.6, 0.7, and 0.8. The calculation is performed for 100,000 time steps for each case. The results are shown in Fig. 8. The pressure differences uniformly increase with increasing inverse droplet radius. The steady slope is equal to the surface tension, which is in excellent agreement with that calculated by the integral method for every EOS and temperature.

**5.4. Suppressed spurious current**

Spurious current is small but finite amplitude circulating flow in the vicinity of a liquid–gas interface with nonzero curvature that occurs in some numerical multiphase models [31]. The above system in which a liquid drop is surrounded by gas can also be used to investigate the spurious current of the present model. When the system evolves to the mechanical equilibrium state, the largest macroscopic velocity in the flow field represents the magnitude of the spurious current. Ideally, the force produced by the surface tension points toward the drop center along the normal direction of the drop surface. Because of the errors of the gradient calculations, the resulting force slightly deviates from the normal direction and causes nonphysical currents. The spurious currents are shown in Fig. 9. Lower temperature leads to higher spurious current (Fig. 9(a)). Because the density ratio is larger at lower temperature, the transition region is steeper. This results in more errors in the gradient calculations and produces higher spurious currents. However, even for temperature of 0.4, the spurious currents of the PR and RKS EOSs are much lower than 0.001 and far better than the previous model [44]. Notably, their density ratios have been greater than 65,000 at temperature of 0.4 (Fig. 9(b)).

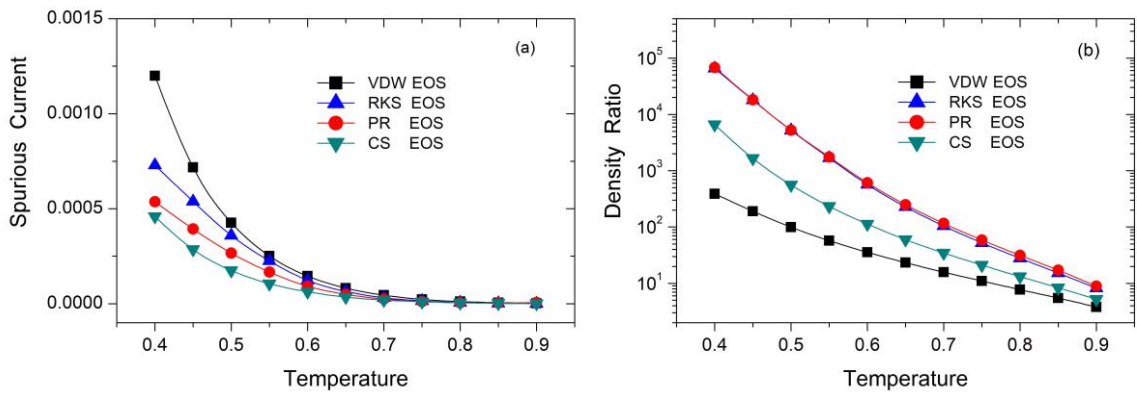

*Fig. 9. (a) Spurious current and (b) density ratio as a function of the temperature. The spurious currents are suppressed to a very low magnitude, even when the density ratios are*

*up to tens of thousands.*

## 5.5. Galilean invariance of a dynamic fluid

A drop splashing on a planar surface with a thin liquid film is simulated as a dynamic test case [40,62]. The PR EOS for water is applied to the liquid–gas system. The computational domain is a rectangular flow field with a width of 1000 lattice units and a height of 300 lattice units. The droplet diameter is $D = 100$ lattice units and the initial droplet speed is $U = 0.1$ down to the surface. The thickness of the liquid film is one-tenth of the height of the flow field. The temperature is chosen to be $T_r = 0.6$ so that the density ratio of the liquid–gas system is close to the ratio of water to vapor. The Reynolds number is defined as $Re = UD/\nu$. The nondimensional time is defined as $t^* = Ut/D$. Periodic boundary condition is applied on the left and right sides, while the chemical-potential boundary condition is adopted on the top and bottom boundaries, where the chemical potential takes the value of the neighbor fluid node [45].

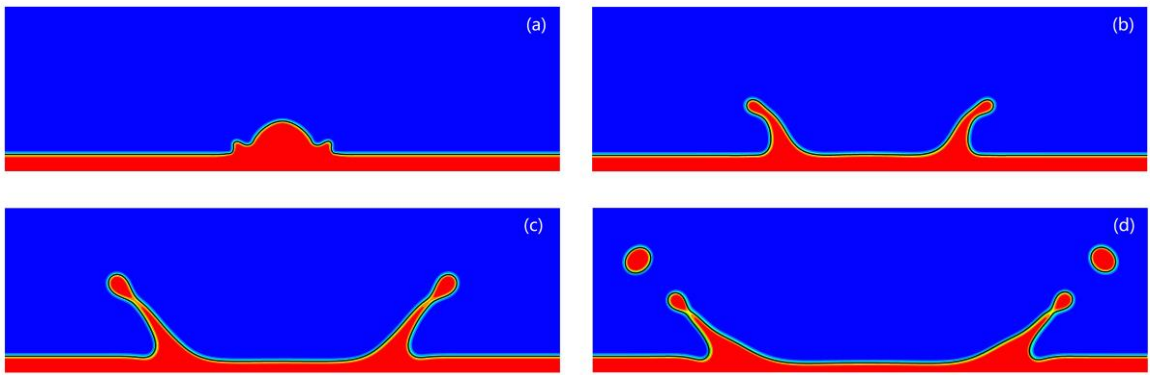

*Fig. 10. Snapshots of the drop impacting a thin liquid film at Re = 150: (a) t\* = 0.5, (b) t\* = 2.0, (c) t\* = 4.5, and (d) t\* = 7.6.*

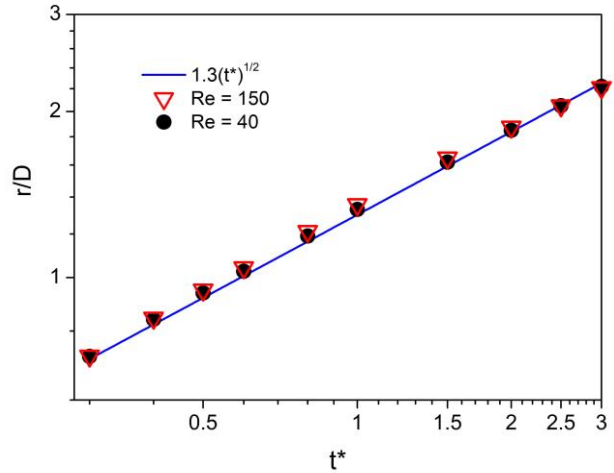

*Fig. 11. Spreading radius of the drop at Re =40 and Re = 150 as a function of the nondimensional time.*

Two drop impact cases are simulated by the present chemical-potential model with Reynolds numbers of 40 and 150. The impact processes for *Re* = 150 are shown in Fig. 10. The impact drop forms a thin liquid sheet at the intersection between the droplet and the liquid film. The liquid sheet grows up tilting upward and outward. It then becomes unstable and the liquid gathers at its top end. Eventually, the sheet breaks up and forms secondary droplets. In contrast, the impact drop of *Re* = 40 does not result in splashing and the droplet motion transforms into a surface wave spreading outward. These observations are consistent with previous studies [40,43]. These studies also found that the spread radius $r$ obeys the power law $r/D \approx C\sqrt{Ut/D}$ in a short time after the droplet impact. The constant *C* is about 1.1 for a three-dimensional model [62] and 1.3 for a two-dimensional model [40]. The spreading radius of the drop as a function of the nondimensional time for *Re* = 40 and 150 is shown in Fig. 11. The present simulations are highly consistent with the power law.

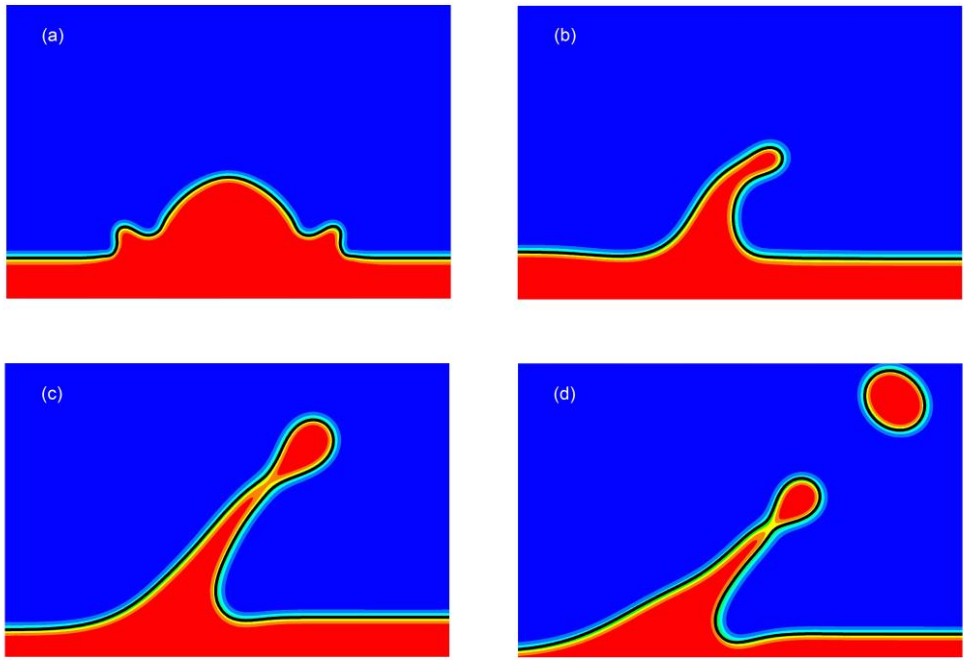

*Fig. 12. Local enlarged images of the drop impact process at Re = 150: (a) t\* = 0.5, (b) t\* = 2.0, (c) t\* = 4.5, and (d) t\* = 7.6. The contours are calculated by simulations with a static reference frame, while the black outlines are obtained from the simulations with a reference speed of 0.03.*

The dynamic case is also applied to verify Galilean invariance of the present method. The simulations are performed relative to different reference frames, whose speed is set to 0 and 0.03 lattice units per time step pointing horizontally to the right. Local enlarged images of the impact process for *Re* = 150 are shown in Fig. 12. The contours are calculated by the simulations with the static reference frame, while the black outlines are obtained from the simulations relative to the moving reference frame. They are in excellent agreement. This verifies that the present model satisfies Galilean invariance in simulation of a dynamic fluid.

### 5.6. Improvement of the stability

Reynolds number is an important indicator in simulation of drop splashing. Usually, small relaxation times are used to reduce the kinematic viscosity and obtain high Reynolds numbers. However, when Reynolds number reaches the level of hundreds, the simulations are prone to be unstable. Here, we attempt to improve the numerical stability of the present model in the context of the chemical-potential model with MRT.

In our cases, the crash always occurs shortly after the simulation begins. It appears that the calculation of chemical potential overflows. After careful analysis, we find that the problem originates from initialization of the numerical simulations. The hyperbolic tangent function in Eq. (41) is a popular scheme to initialize the density field of a liquid–gas coexistence system [38,40]. It provides a smooth transition region between the liquid phase and gas phase. However, as mentioned in Section 3.3, the real profile of a transition region is different from a curve of a hyperbolic tangent function. On the other hand, the densities of the liquid and gas used in Eq. (41) are calculated by the Maxwell equal-area construction in a one-dimensional system with mechanical equilibrium. The initial pressure inside the drop is the same as that outside. Both are different from the pressures indicated by the Young–Laplace equation. Once the simulation starts, these differences are quickly adjusted and form some waves of pressure, which will converge at the center part of the drop after hundreds of time steps. Because of the coupling of the density and pressure in LBM, the convergences result in violent fluctuations of the density. Taking the PR EOS as an example, when the density is greater than or equal to $b^{-1}$, the first term of the right side of Eq. (19) overflows and the simulation crashes. The change of the chemical potential with the density is shown in Fig. 13, where the dashed line represents the position of $b^{-1} = 21/2$. It is clear that even when the density is less than $b^{-1}$, the chemical potential rapidly increases when it approaches $b^{-1}$. This could lead to a very large gradient of chemical potential and then, according to Eq. (12), produce an abnormal nonideal force, which is too large for the simulation to continue running.

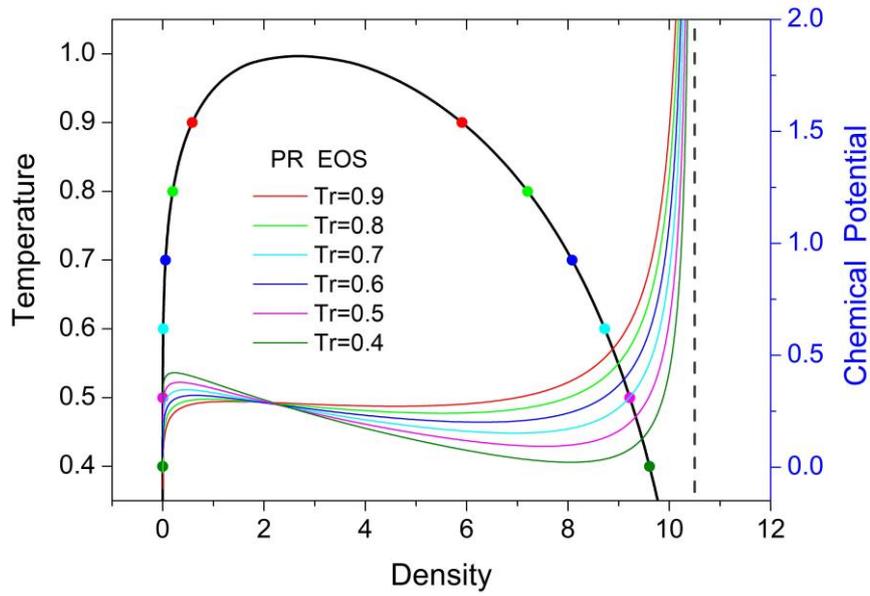

*Fig. 13. Liquid–gas coexistence densities calculated by the Maxwell equal-area construction and chemical potential at a given temperature changed with the density fluctuations.*

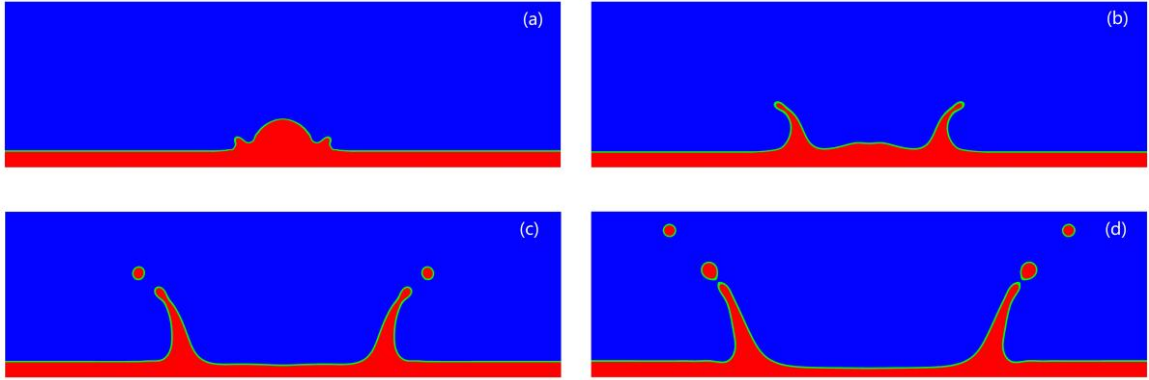

*Fig. 14. Snapshots of the drop impacting a thin liquid film at Re = 400: (a) t\* = 0.5, (b) t\* = 1.5, (c) t\* = 3.0, and (d) t\* = 4.5.*

Based on the above analysis, the density fluctuations and resulting crashes originate from some nonphysical factors. Here, a simple scheme is proposed to improve the stability. An upper limit is applied to control the chemical potential to be not too large. It only works in the first few hundred time steps of the simulation. For the present simulations of drop splashing, the upper limit is 6 for the PR EOS with $Tr = 0.6$. The number may be different for other EOSs and temperatures. This scheme effectively increases the Reynolds number. The impact processes for *Re* = 400 are shown in Fig. 14. The impact drop makes a thinner liquid sheet at the intersection between the droplet and the liquid film. The liquid sheet more rapidly grows upward and outward. The sheet then breaks up and forms secondary droplets. It should be stressed that the scheme only limits the chemical potential and does not affect the density or distribution functions. Therefore, the scheme does not damage the mass and momentum conservation of the system.

## 6. Conclusions

The errors in nonideal force evaluations are the main reason why some multiphase models [35,44] can only simulate a liquid–gas density ratio of about 100. In this study, we improve the accuracy of the nonideal force and achieve a chemical-potential multiphase LBM with extremely large density ratios. A proportional coefficient is introduced to connect the mesh step to the lattice step, so the mesh space is decoupled from the momentum space. Owing to the smaller mesh step, the transition region is described by many more nodes. That is, the steep transition region in the momentum space is stretched into a gentler curve in the mesh space. The widely used CDM has low accuracy and damages the stability of multiphase models. CFM is applied to calculate the gradients in the present multiphase model to replace CDM. The scheme is formally sixth-order accurate, so the gradients of the density and chemical potential obtain very high accuracy. Because a dimensional transformation relates the mesh space to the momentum space and there is no loss of accuracy in the transformation process, the present model is mathematically equivalent to previous versions [9,44], which have been confirmed to theoretically satisfy the thermodynamic consistency and Galilean invariance. Numerical simulations verify the theoretical analysis. The resulting two-phase coexistence densities are in excellent agreement with the predictions of the Maxwell equal-area construction until very low temperatures. The liquid–gas density ratios reach more than $10^{14}$. Moreover, simulations of drop splashing confirm that the present model is Galilean invariant for a dynamic flow field.

Different proportional coefficients define different mesh spaces; although these mesh spaces connect to the same momentum space. With a series of proportional coefficients, the interface width and surface tension are calculated. They proportionally change with the coefficient and have the same values when they are transformed to the momentum space. This indicates that the transformations between the momentum space and computational mesh are stable and accurate. Owing to the high-accuracy nonideal force evaluations, the spurious currents are suppressed to a very low level, even though the density ratio reaches tens of thousands. The Young–Laplace equation is used to verify the surface tension for various popular EOSs. An upper limit of chemical potential is used to improve the stability of the dynamic simulations. The present model is implemented in the SRT and MRT models. For the simulations of static flow fields, the results are difficult to distinguish. For dynamic drop splashing, the MRT model shows better stability. This study greatly extends the applications of the chemical-potential multiphase model, and these techniques can also be used to improve other multiphase models. In future work, the present model will be comprehensively compared with other multiphase models, especially the pseudopotential model. In addition, more efforts will be devoted to improve the stability and achieve higher Reynolds number.

**Acknowledgments**

This work was supported by the National Natural Science Foundation of China (Grant Nos. 11862003, 81860635, 91741101, 91752204), the Key Project of Guangxi Natural Science Foundation (Grant No. 2017GXNSFDA198038), Guangxi "Bagui Scholar" Teams for Innovation and Research Project, Guangxi Collaborative Innovation Center of Multi-source Information Integration and Intelligent Processing.